\documentclass[aps,prd,12point,twocolumn,nofootinbib,showpacs,superscriptaddress]{revtex4-1}
\def\theequation{\arabic{section}.\arabic{equation}}
\usepackage{psfrag}
\usepackage{subfigure}
\usepackage{color}
\usepackage{mathrsfs}
\usepackage{graphicx}
\usepackage{amssymb, bm}
\usepackage{amsmath, amsthm}
\usepackage{epstopdf}
\usepackage{hyperref}
\usepackage{enumerate}
\usepackage{longtable}

\newcommand{\be}{\begin{equation}}
\newcommand{\ee}{\end{equation}}

\newcommand{\rc}{\color{red}}

\begin{document}
\def\theequation{\arabic{section}.\arabic{equation}} 

\title{Turning a Newtonian analogy for FLRW cosmology into a relativistic 
problem}

\author{Valerio Faraoni}
\email[]{vfaraoni@ubishops.ca}
\affiliation{Department of Physics and Astronomy, Bishop's University, 
2600 College Street, Sherbrooke, Qu\'ebec, 
Canada J1M~1Z7}

\author{Farah Atieh}
\email[]{fatieh19@ubishops.ca}
\affiliation{Department of Physics and Astronomy, Bishop's University, 
2600 College Street, Sherbrooke, Qu\'ebec, 
Canada J1M~1Z7}



\begin{abstract}

A Newtonian uniform ball expanding in empty space constitutes a common 
heuristic analogy for FLRW cosmology. We discuss possible implementations 
of the corresponding general-relativistic problem and a variety of new 
cosmological analogies arising from them.  We highlight essential 
ingredients of the Newtonian analogy, including that the quasilocal energy 
is always ``Newtonian'' in the sense that the magnetic part of the Weyl 
tensor does not contribute to it. A symmetry of the Einstein-Friedmann 
equations produces another one in the original Newtonian system.

\end{abstract}


\maketitle

\section{Introduction}
\setcounter{equation}{0}
\label{sec:1}
	
A Newtonian analogy is often used to introduce 
Friedmann-Lema\^itre-Robertson-Walker (FLRW)  cosmology to beginners and 
provide physical intuition ({\em e.g.}, \cite{Ryden, Liddle, Inverno, 
Slava}---see \cite{Harrison} for a different version). The analogy is 
based on a ball of uniform density expanding in 
empty space and on conservation of energy for a test particle on the 
surface of this ball. The Newtonian energy conservation equation is 
formally analogous to the Friedmann equation of relativistic cosmology for 
a universe filled with a perfect fluid with zero pressure (a dust). Of 
course, this 
analogy is not realistic and the proper description of the universe 
requires general 
relativity (GR). Moreover, the analogy has limitations because the 
Newtonian ball 
can only produce the analogue of a dust-filled universe, while FLRW 
cosmology includes a much richer variety of matter content (radiation, 
dark energy, scalar fields, {\em etc.}). Most presentations in the 
literature and unpublished lecture notes available on the internet do 
caution that this is only an analogy.  {\em A posteriori},  it is 
interesting to revisit this analogy from the GR point of view. Does the 
corresponding GR problem still lead to an analogy with FLRW cosmology? 
Does a relativistic isolated ball still expand like a FLRW universe? Like 
a dust-dominated universe, or are there other possibilities?  A similar 
situation involves the black hole concept, first 
introduced by Michell and Laplace in a naive Newtonian context 
\cite{Michell,Laplace}, and then rediscovered in the Schwarzschild 
solution of the Einstein equations.

Here we revisit the Newtonian analogy by considering an expanding (or 
contracting) isolated ball of uniform density in empty space in GR, and 
two possible ways to extend the Newtonian analogy. The 
first, and easy, way consists of cheating the difficulties and looking 
at the radial motion of a test particle just outside the surface of 
the 
ball,  {\em i.e.}, using the radial timelike geodesic equation in the 
vacuum, 
spherical geometry (which is necessarily Schwarzschild). The radial 
timelike geodesic 
equation is still formally analogous to the Friedmann equation for a 
dust-filled universe. For completeness, we consider also radial null  
geodesics, for which the analogous Friedmann equation provides the 
empty Milne universe. This ``easy'' way to derive an analogy has 
significant 
limitations, which are discussed below. However, it can be generalized to 
many static and spherically symmetric geometries, producing  a host of new 
cosmological analogies.

The second, and proper, way to address the problem consists of looking for 
exact solutions of the Einstein equation that are spherically symmetric, 
time-dependent, and asymptotically flat and are sourced by a ball of 
perfect fluid with uniform density and pressure, expanding or contracting 
{\em in vacuo}. Due to the Birkhoff theorem, the solution outside the ball 
is the Schwarzschild geometry while the interior solution, to be 
determined, must be matched to it on the surface of the ball by imposing 
the Darmois-Israel junction conditions \cite{Lanczos, Darmois, Israel}. 
The interior of the ball will necessarily be described by a FLRW solution 
(hence this system would be useless to introduce FLRW cosmology, but this 
is no longer the motivation).

The solution of this GR problem can be obtained as a special case of exact   
solutions of the Einstein equations describing spherical 
objects embedded in FLRW spaces or external fluids. A   
well-known one corresponds to the 
Oppenheimer-Snyder solution for dust collapse  
\cite{OppenheimerSnyder}, in which the  ball interior is 
positively curved FLRW. This is, however, only one 
possibility and one would like to consider expanding spheres, possibly 
with a range of equations of state. 
The situation resembles that of an expanding fireball or spherical 
explosion, and there are already in the 
literature analytical solutions of the Einstein equations that can be used 
to solve  our problem.

Assuming the FLRW line element
\be
ds^2=-dt^2+a^2(t) \left( \frac{dr^2}{1-Kr^2} +r^2 d\Omega_{(2)}^2 \right) 
\ee
in comoving coordinates $\left( t,r, \vartheta, \varphi \right)$, where 
the 
constant $K $ is the curvature index and $d\Omega_{(2)}^2=d\vartheta^2 
+\sin^2 \vartheta \, d\varphi^2$ is the line element on the unit 2-sphere, 
the Einstein-Friedmann equations for a universe sourced by a perfect fluid 
with stress-energy tensor $T_{ab}=\left(P+\rho \right) u_a u_b +Pg_{ab} $ 
(where $\rho$ is the energy density, $P$ is the pressure, and $u^c$ is the 
4-velocity of comoving observers) are\footnote{We follow the notation of 
Ref.~\cite{Wald} and use units in which the speed of light and Newton's 
constant $G$ are unity.} 
\begin{eqnarray}
&& H^2 \equiv \left( \frac{\dot{a}}{a} \right)^2 = \frac{8\pi}{3} \, 
\rho -\frac{K}{a^2} \,,\label{FFrr}\\
&&\nonumber\\
&&\frac{\ddot{a}}{a} = -\frac{4\pi}{3} \left( \rho+3P \right) \,,\\
&&\nonumber\\
&&\dot{\rho}+3H \left( P+\rho \right) =0 \,,\label{conservation}
\end{eqnarray}
where an overdot denotes differentiation with respect to the comoving 
time $t$ and $H\equiv \dot{a}/a$ is the Hubble function. Only two of these 
three equations are independent.

We proceed as follows: the next section recalls the Newtonian analogy; 
Sec.~\ref{sec:3} points out the analogy between radial timelike/null 
geodesics of Schwarzschild and the Friedmann equation and generalizes it 
to many spherical static geometries. Sec~\ref{sec:4} 
discusses the proper GR problem, highlighting essential features of the 
Newtonian analog, while Sec.~\ref{sec:5} discusses the Newtonian character 
of the quasilocal mass used. Sec.~\ref{sec:6} uses a little known symmetry 
of the Einstein-Friedmann equations to generate (only for zero 
energy/spatially flat universes), a new symmetry of the Newtonian ball, 
while Sec.~\ref{sec:7} contains the conclusions.

\section{Newtonian analogy for FLRW universes}
\label{sec:2}
\setcounter{equation}{0}

Consider, in Newtonian physics,\footnote{We restore Newton's constant $G$ 
and the speed of light $c$ in this section.} a uniform expanding  
sphere with radius $R(t)$, homogeneous 
density $\rho(t) $, and 
total mass $M=4\pi R^3 \, \rho/3 $, and   a test particle of 
mass $m$ on its surface. The total mechanical energy of 
this particle is
\be
E=\frac{1}{2} \, m \dot{R}^2 -\frac{GMm}{R}  \label{energy}
\ee
and is constant. Re-arranging this energy integral, we write
\be
\frac{\dot{R}^2}{R^2}= \frac{8\pi }{3} \, \rho +\frac{2E}{mR^2} 
\,.\label{noi}
\ee
By introducing the quantities
\be
H \equiv \frac{ \dot{R}}{R} \,, \;\;\;\;\;\; K \equiv - \, \frac{2E}{mc^2} 
\,,\label{jokeFriedmann}
\ee
where $c$ is the speed of light, Eq.~(\ref{noi}) becomes
\be
H^2 =\frac{8\pi }{3} \, \rho -\frac{Kc^2}{R^2} \,.
\ee
This equation is analogous to the Friedmann 
equation of relativistic cosmology for a universe filled by 
non-relativistic matter.

If $E>0$, which corresponds to $K<0$ and to $v>v_\text{escape}$, where 
\be
v_\text{escape}=\sqrt{ \frac{2GM}{R}}
\ee
is the escape velocity from the surface of the ball, then the particle 
will escape to $R=+\infty$ with residual velocity $\dot{R}_{\infty}$ 
given by the limit
\be
0< E=\frac{1}{2} \, m \dot{R}^2 -\frac{GMm}{R} \rightarrow \frac{1}{2} 
\, 
m \dot{R}_{\infty}^2  \,.
\ee

If $E=0$, corresponding to $K=0$ and to $v=v_\text{escape}$, the 
particle 
barely escapes to infinity with zero velocity $\dot{R}$, according to  
\be
0=\frac{1}{2} \, m \dot{R}^2 -\frac{GMm}{R} \rightarrow \frac{1}{2} \, 
m \dot{R}_{\infty}^2  =0 \,.
\ee

If instead $E<0$, corresponding to $K>0$ and to $v<v_\text{escape}$, 
the 
particle reaches a maximum radius and then falls back reversing its 
velocity. In this case, one cannot take the limit $R\rightarrow +\infty$.  
The maximum radius is attained 
when $\dot{R}=0$ just before the particle reverses its 
velocity:
\be
E= -\frac{GMm}{R_{max}} \,,
\ee
which yields 
\be
R_{max}= \frac{GMm}{|E|} \,.
\ee

Analytical solutions of the energy 
integral~(\ref{jokeFriedmann}) in parametric form are well known.  Define 
the new time variable $\eta$ by 
\be
d\eta= \sqrt{ \frac{2\left| E \right|}{m} } \, 
\frac{dt}{R(t)} \,;
\ee
then the solutions are
\begin{eqnarray}
&&R= \frac{GMm}{2\left| E \right|} \left( 1-\cos\eta  
\right)\,, \\
&& \nonumber\\
&& \pm \left( t-t_0 \right) \sqrt{ \frac{2\left| E 
\right|}{m}}= 
\frac{GMm}{2\left| E \right|} \left( \eta -\sin \eta  
\right)
\end{eqnarray}
if $E<0$, or
\be
R(t) \propto \left( t-t_0 \right)^{2/3} 
\ee
if $E=0$ (in this case one can eliminate the parameter), or
\begin{eqnarray}
&& R= \frac{GMm}{2\left| E \right|} \left( \cosh\eta -1 
\right)\,, \\
&& \nonumber\\
&& \pm\left( t-t_0 \right) \sqrt{ \frac{2 E }{m}} = 
\frac{GMm}{2 E} \left( \sinh \eta -\eta \right)\,, 
\end{eqnarray}
if $E>0$.

Because the energy $E$ is conserved, it is not possible for a closed 
universe to become an open ({\em i.e.}, 
expanding forever) one and {\em vice-versa}, or for a closed or  
open universe to become  flat  and {\it vice-versa}.

In this Newtonian analogy for the universe, the solution for $E<0$ 
corresponds to an open universe that expands forever, the case $E>0$ to a 
universe that reaches a maximum size and recollapses, and the $E=0$ case 
to a critically open universe. This is only an analogy: in relativistic 
cosmology the meaning 
of the variables is different and there is no centre of expansion for the 
universe. Nevertheless, this analogy recurs frequently in the literature 
and is even used as a toy model for quantum cosmology 
\cite{Vieira1,Vieira2}.

\section{Test particle starting above the ball surface}
\setcounter{equation}{0}
\label{sec:3}
	
Now let us promote the Newtonian problem to a general-relativistic one. We 
must derive a new, relativistic equation for the motion of the surface of 
a uniform ball of fluid, which is a non-trivial task best postponed to the 
next section. Here we consider a simpler alternative: in the Newtonian 
analogy, things proceed unchanged if the test particle is just above the 
surface of the ball instead of being located on it, and it shoots radially 
away from this surface. Things are unchanged as long as the expanding ball 
does not overtake the particle, or the falling particle returning from a 
failed escape hits the ball (whether this happens depends on how the 
uniform ball expands, which in turn depends on the nature of the fluid in 
it, the equation of motion it satisfies, and the initial velocities of 
test particle and ball surface). That this does not happen is certainly 
not warranted {\em a priori} and will be discussed in the following 
sections. For now, let us proceed by assuming that test particle and ball 
do not meet, at least for a certain period of time.

In GR, nobody forbids to consider a particle {\em outside} the ball, that 
starts out radially close to it, as long as the surface of the ball does 
not overtake the particle. For a contracting ball this does not happen at 
least until the particle reaches its maximimum height and comes back (if 
it does). If it comes back, it would have to fall radially faster than the 
ball contracts, which is possible if the matter in the ball has pressure 
and the fluid does not follow geodesics. Or, a falling particle could hit 
an expanding ball.  It is also possible that the surface of the expanding 
ball moves outward faster than the massive test particle, overtaking it. 
This will, again, happen if the fluid has pressure, or if its initial 
velocity is higher than that of the particle, {\em etc.} Separating the 
test particle from the surface of the ball opens up these new 
possibilities.

Assuming, for the moment, that the test particle and the ball surface do 
not collide, our massive test particle moves along a radial timelike 
geodesic of the Schwarzschild geometry
\be
ds^2=-\left(1-\frac{2m}{\bar{r}}\right)dt^2 
+\frac{d\bar{r}^2}{1-\frac{2m}{\bar{r}}} + \bar{r}^2 
d\Omega_{(2)}^2 \,. \label{Schwarzschild}
\ee
 The equation of radial geodesics is \cite{Wald} 
\be 
\dot{\bar{r}}^2 +\left( 
1-\frac{2m}{\bar{r}}\right) \left( \frac{L^2}{\bar{r}^2} +\kappa \right) 
=E^2 
\label{eq:geodesic} 
\ee 
for $\bar{r}>2m$, where an overdot now denotes differentiation with 
respect to the proper time $\tau$ or the affine parameter along the 
geodesic, and $E$ and $L$ are 
the conserved energy and angular 
momentum per unit mass of the 
particle, which are (apart from the sign) contractions of the particle 
4-velocity $u^a$ with the timelike and rotational Killing vector 
fields, respectively. In Eq.~(\ref{eq:geodesic}), $\kappa=1$ for timelike 
geodesics and $\kappa=0$ for null geodesics.

\subsection{Massive test particle}
\label{subsec:2.1}

In the timelike case, the radial geodesic equation can be written as
\be
\left( \frac{\dot{\bar{r}}}{\bar{r}} \right)^2 = \frac{E^2-1}{\bar{r}^2} 
+\frac{2m}{\bar{r}^3}\,. 
\ee
This equation is analogous to the Friedmann equation~(\ref{FFrr}) for a 
FLRW universe
where $K=1-E^2$ plays the role of  the curvature index, $\rho=\rho_0/a^3$ 
corresponds to a 
radiation fluid  with equation of state $P=\rho/3$, and $\rho_0= 
\frac{3m}{4\pi } $. Therefore, there is a straightforward analogy between 
the motion of a test particle in the field of the ball and the Friedmann 
equation, as in the 
Newtonian situation. All three possible signs of the curvature $K$ of the 
analogous FLRW universe are possible, but only an analogous universe 
containing a dust fluid can be obtained.

\subsection{Massless test particle}
\label{subsec:1.2}

We now turn to radial null geodesics, a possibility that   
does not exist in Newtonian physics. In this case, the ball 
surface will never overtake, or reach, a photon starting radially above 
it, which always escapes to $\bar{r}=+\infty$. The photon trajectory 
satisfies 
\be
\dot{\bar{r}}^2=E^2 \,,
\ee
where the overdot now denotes differentiation with respect to an affine 
parameter along the null geodesic. The cosmological analogue of this 
trajectory  is 
\be
H^2=-\frac{K}{a^2}
\ee
with $K<0$. This is the Milne universe, {\em i.e.}, empty Minkowski space 
in accelerated  coordinates, sliced using a hyperbolic  
foliation  
({\em e.g.}, \cite{Slava}). The 
scale factor solving the Friedmann equation is now linear, $a(t)=t 
$, and all components of the Riemann tensor vanish. Writing the line 
element in hyperspherical coordinates as
\be
ds^2=-dt^2+t^2 \left( d\chi^2 +\sinh^2 \chi \, d\Omega_{(2)}^2 \right) \,,
\ee
the coordinate 
transformation $\tau = t\cosh \chi \,, r=t\sinh \chi$ then  
brings the FLRW line element into the Minkowski form $ds^2=-d\tau^2 +dr^2 
+r^2 d\Omega_{(2)}^2$.

\subsection{Generalization to any static spherical geometry}

As a digression, we note that the cosmological analogy for timelike 
and null geodesics can often be generalized  to static and spherically 
symmetric geometries. For such spacetimes, the line element can always be 
written in the Abreu-Nielsen-Visser gauge \cite{NielsenVisser,AbreuVisser} 
as
\begin{equation}
ds^2 = -\mbox{e}^{-2 \Phi (R)} \left(1 -  \frac{2M(R) }{R}\right) dt^2 + 
\frac{dR^2}{1-\frac{2M(R)}{R}} + R^2 d\Omega_{(2)} ^2 
\end{equation}	
where $ T^a \equiv \left ( \partial/\partial t \right)^a $
is the timelike Killing vector and $M(R) $ is the 
Misner-Sharp-Hernandez mass \cite{MSH1, MSH2} (to which the 
Hawking-Hayward quasilocal mass \cite{Hawking, Hayward} reduces in 
spherical symmetry \cite{Haywardspherical}). Consider radial timelike 
geodesics: the energy $E$ of a particle of mass $m$ and 4-momentum 
$p^c=mu^c $ along such a geodesic is conserved, $
p_c T^c = -E $,
which yields 
\begin{equation}
u^0 = \frac{dt}{d\tau} = \frac{\bar{E}\, \mbox{e}^{2 \Phi} }{1 -  
\frac{2M}{R}} \,, \label{xxx}
\end{equation}
where $\bar{E} = E/m$ is the energy per unit mass.  The normalization $u_c 
u^c =-1 
$ gives 
\be 
g_{00} \left( \frac{dt}{d\tau} \right)^2 + g_{11} 
\left( \frac{dR}{d\tau} \right)^2 = -1 \,;
\end{equation}
substituting Eq.~(\ref{xxx}) to obtain $\left( dR/d\tau\right)^2$ and 
dividing the result by $R^2$, one obtains 
\begin{equation}
\left( \frac{1}{R} \, \frac{dR}{d\tau} \right)^2 = \frac{\bar{E}^2 \, 
\mbox{e}^{2 \Phi}}{R^2} - \frac{1}{R^2} + \frac{2 M(R)}{R^3} 
\,.\label{pseudoFE}
\end{equation}

Many spherical spacetimes of interest in GR and in alternative theories of 
gravity satisfy the condition  $g_{tt} g_{RR} =-1$  (or $ \Phi \equiv 
0$), which is associated to special algebraic properties \cite{Jacobson}; 
in this case Eq.~(\ref{pseudoFE}) reduces to
\begin{equation}
\left( \frac{1}{R} \, \frac{dR}{d\tau} \right)^2 = \frac{\left( \bar{E}^2 
-1\right)}{R^2}  + \frac{2 M(R)}{R^3}\,,
\end{equation}
which can be analogous to the Friedmann equation~(\ref{FFrr}) for a 
universe with curvature index $K = 1-\bar{E}^2$ and energy 
density $\rho = \frac{3}{4\pi } \, \frac{M(a)}{a^3}$. To complete 
the analogy, the cosmic fluid must satisfy the covariant conservation 
equation~(\ref{conservation}), which yields
\begin{equation}
P= - \rho - \frac{a}{3} \, \frac{d\rho}{da} = - \rho - \frac{a}{4 \pi } 
\left(\frac{ M'}{a^3} - \frac{3M}{a^4}\right) = -\frac{M'}{4 \pi a^2} \,.
\end{equation}
This effective equation of state can be written in the time-dependent form 
\begin{equation}
P =-\frac{M'(a)}{4\pi a^2} \equiv w(a)  \rho \,.
\end{equation}

Let us turn now to radial null geodesics. The energy $E$ of a photon is 
conserved along each such geodesic, $u_c T^c=-E$, giving
\begin{equation}
\frac{dt}{d\lambda} = \frac{E \, \mbox{e}^{2 \Phi}}{1-  
2M(R)/ R} \,,
\ee
where $\lambda$ is an affine parameter. Substitution into the 
normalization $ u_c u^c=0$ yields
\begin{equation}
\left(\frac{1}{R} \, \frac{dR}{d\lambda}\right)^2 = 
\frac{E^2  \mbox{e}^{2\Phi(R)}}{R^2}
\end{equation}
If $\Phi \equiv 0$ this equation is analogous to the Friedmann 
equation $ H^2 = - K/a^2 $, where $K= -E^2 < 0$, producing Minkowski space 
disguised as the Milne universe. If $\Phi\neq 0$, there is a chance of  
a more meaningful analogy. Examples of cosmic analogies derived from 
radial timelike and 
null geodesics of static spherical geometries are listed in the Appendix.

\section{Exact GR solution for an expanding relativistic ball}
\setcounter{equation}{0}
\label{sec:4}
	
Now let the massive particle sit on the surface of the ball and be a 
particle of the fluid composing the ball, which is always larger than its 
Schwarzschild radius. For a general fluid, this particle does 
not follow a timelike geodesic. In fact, in the presence of pressure, the 
4-gradient $\nabla_a P$ moves fluid particles away from 
geodesics. Only dust ($P= 0$) is geodesic in the absence of 
external forces.

Since we cannot ignore the matter at radii below the initial particle 
radius $R(0)$, we must now find, and solve, the equation of motion  
for the boundary of the relativistic ball.

A static ball with uniform density is described by the well 
known Schwarzschild 
interior solution \cite{Tolman}, but it is of no use here. We want instead 
a ball of 
uniform density that expands/contracts while remaining uniform. The metric 
must be asymptotically flat (we assume zero 
cosmological constant). Due to Birkhoff's theorem, an observer outside the 
matter distribution (or on the ball surface) sees the Schwarzschild 
vacuum. 
What is the equation of the surface of the ball in this case? Is this 
motion geodesic or does the normal to the ball surface deviate from a 
geodesic vector?  
Moving on to time-dependent, asymptotically, flat fluid spheres, some 
exact 
solutions were provided early on by Vaidya \cite{Vaidyaball}. They contain 
the Schwarzschild interior solution \cite{Tolman} and the 
Oppenheimer-Snyder \cite{OppenheimerSnyder} solution as special cases. The 
most useful geometries here are probably those of Smoller and Temple 
\cite{SmollerTemple}, which contain the solution of the Einstein equations 
describing our situation as a special case and reproduce results of 
previous literature, that we briefly review here.

Mashhoon and Partovi \cite{MashhoonPartovi} found that the unique solution 
for the spherically and shear-free motion of an uncharged perfect fluid 
obeying an equation of state is the FLRW one. However, another 
hypothesis must be added for the theorem to hold, namely that the energy 
density is uniform, $\partial \rho/\partial r=0$, which is one of our 
needed assumptions \cite{SrivastavaPrasad}. Therefore, the interior of the 
ball can only be FLRW. {\em A posteriori}, this fact 
legitimates the use of a uniform ball in the Newtonian analogy. A 
non-uniform ball would lead to a spherically symmetric, but 
inhomogeneous, universe. Contracting balls with pressure were considered 
by Thompson and Whitrow \cite{ThompsonWhitrow, ThompsonWhitrow2} and Bondi 
\cite{Bondi,BondiNature}, mainly to study gravitational 
collapse to black holes.

The Newtonian analogy requires also that the particles composing the ball, 
as well as the test particle on its surface, have no pressure and are 
subject only to gravity: the fluid composing the ball must be 
a dust for the interior geometry to match the exterior 
Schwarzschild one.

\subsection{A special case of the Smoller-Temple shock wave solution} 
\label{subsec:3.1}

Smoller and Temple \cite{SmollerTemple} consider exact solutions of the 
Einstein equations 
representing a spherical shock wave expanding into a gas, with interior 
and exterior matching on a zero thickness shell with no material on it (no 
jump). Because of uniformity, $\rho=\rho(t), P=P(t)$, the metric inside 
the shock wave must be FLRW (all values of the curvature 
index $K$ are possible),
\be
ds^2=-dt^2 +a^2(t) \left( \frac{dr^2}{1-Kr^2} +r^2 d\Omega_{(2)}^2 
\right) \,.
\ee
Outside the shock wave, the geometry is that of a static and spherical  
Oppenheimer-Tolman solution (usually employed to 
describe the interior of a relativistic  star)
\be
d\bar{s}^2= -B(\bar{r})dt^2 + 
\frac{d\bar{r}^2}{1-2M/\bar{r}}+\bar{r}^2 d\Omega_{(2)}^2 
\label{OTinterior}
\ee
in coordinates\footnote{The comoving coordinate $ r $ inside the ball is 
not the same as the curvature coordinate $  \bar{r}$ outside.} 
$\left( t, \bar{r}, \vartheta, \varphi \right)$, where 
$M(\bar{r})$ is the mass contained inside the sphere 
of radius 
$\bar{r}$, or
\be
\frac{dM}{d\bar{r}}=4\pi \bar{r}^2 \rho(\bar{r}) \,,\label{OTmass}
\ee
$ \rho(\bar{r})$ is the energy density at radius $\bar{r}$, and 
\be
\frac{B'(\bar{r})}{B(\bar{r})} = -\, 
\frac{2\bar{P}'(\bar{r})}{\bar{\rho}+ \bar{P} }\,.
\ee
$\bar{P}$ is the outside pressure, and a prime denotes differentiation 
with respect to this radial coordinate. The mass $M(\bar{r})$ coincides 
with the Misner-Sharp-Hernandez mass \cite{MSH1, MSH2} at radius 
$\bar{r}$, which is defined in any spherically symmetric spacetime by
\be
1-\frac{2M_\text{MSH}}{R} = \nabla^c R \nabla_c R \, \dot{=}g^{RR} \,\,,
\ee
where $R$ is the areal radius and the last equality holds if the areal 
radius is employed as the radial coordinate (which is the case for the 
line element~(\ref{OTinterior})).

The interior and exterior solutions are matched on the surface of a ball 
(the shock wave) by imposing the Darmois-Israel junction conditions 
\cite{Lanczos, Darmois, Israel} so that there is no jump in the first and 
second fundamental forms and there is no material layer on the shell. The 
junction conditions for an outgoing shock wave give 
\cite{SmollerTemple} 
\begin{eqnarray} 
r\dot{a} &=& \sqrt{1-Kr^2} \, \sqrt{ 
1-\Theta} \,,\label{q1}\\ &&\nonumber\\ \dot{r}a &=& \sqrt{1-Kr^2} \, 
\frac{ \sqrt{ 1-\Theta}}{ \gamma\Theta -1}\,,\label{q2} 
\end{eqnarray} 
where 
\begin{eqnarray} 
\Theta &=& \frac{ 1-2M/r}{1-Kr^2} \,,\\ 
&&\nonumber\\ 
\gamma &=& \frac{ \rho+\bar{P}}{\bar{\rho}+\bar{P}} \,. 
\end{eqnarray} 
A coordinate transformation between interior and exterior 
coordinates is found in \cite{SmollerTemple}, with the simple result 
\be 
\bar{r}=a(t)\, r \,, 
\ee 
implying that the areas of 2-spheres of symmetry 
change smoothly across the shock wave ({\em i.e.}, the areal radius 
$R=a(t)r$ of the FLRW interior matches the areal radius $\bar{r}$ of the 
Schwarzschild exterior) and that the surface of the ball comoves with its 
FLRW interior. Now we impose that the exterior is vacuum, 
$\bar{\rho}=\bar{P}=0$, then the spherical and asymptotically flat 
exterior metric necessarily reduces to Schwarzschild. This situation is 
reported as a special case of the outgoing shock wave in 
\cite{SmollerTemple}. It requires that the pressure $P$ vanishes inside 
the entire ball in order to match the vanishing pressure at the boundary: 
the interior FLRW fluid can only be dust.\footnote{A similar situation is 
encountered in Swiss-cheese models \cite{Krasinskibook}.} In the Newtonian 
analogy, one 
wants a particle on the ball surface subject only to gravity: a pressure 
gradient would complicate the Newtonian description, requiring the 
specification of an equation of state, and could make the ball overtake a 
test particle placed on it (these complications would ruin the simplicity 
of the Newtonian analogy).

The vanishing of the outside pressure $\bar{P}\rightarrow 0$ corresponds 
to the limit $\gamma\rightarrow \infty $ and implies \cite{SmollerTemple} 
\be
r=r_0=\mbox{const.}
\ee
at the surface of the ball. Then the density must scale as
\be
\rho(t)= \frac{3M}{4\pi r_0^3 a^3} \,,
\ee
consistent with a dust, and the Friedmann equation at this surface reduces 
to
\be
\dot{a}^2 = \frac{2M}{r_0^3 a} -K \,.
\ee
Apparently unknown to Smoller and Temple, the conclusion that only a 
uniform ball of dust can be matched to Schwarzschild was reached already 
by McVittie \cite{McVittie}, Bondi \cite{Bondi}, Mansouri \cite{Mansouri}, 
Mashhoon and Partovi \cite{MashhoonPartovi}, and Glass \cite{Glass}. The 
surface of the ball expands into the surrounding Schwarzschild vacuum 
while comoving with its interior. Further, since $P=0$, the 
fluid particles follow radial geodesics because of spherical symmetry, and 
the unit normal to the comoving ball surface is a timelike geodesic 
vector (radial geodesic congruences are normal to the surface of the 
ball because, due again to spherical symmetry, there is no vorticity). The 
radial geodesics of the interior geometry join smoothly the radial 
geodesics of the Schwarzschild exterior, provided that the ball surface 
comoves with its interior \cite{SmollerTemple}.

All possibilities for the spatial curvature of the FLRW space inside the 
ball are studied in \cite{SmollerTemple}: for $K>0$ the well-known 
Oppenheimer-Snyder solution describing the collapse of a ball of dust 
\cite{OppenheimerSnyder} is 
recovered by considering an ingoing shock wave (and changing the signs of 
the right-hand sides of Eqs.~(\ref{q1}), (\ref{q2})), with the ball 
boundary describing the cycloid 
\begin{eqnarray}
a(\eta) & = & \frac{1}{2} \left( 1+ \cos \eta \right) \,,\\
&&\nonumber\\
t(\eta) &=& \frac{1}{2\sqrt{K}} \left( \eta+ \sin \eta \right) \,,
\end{eqnarray}
where the initial conditions $a(0)=1, \dot{a}(0)=0$ have been imposed and 
the curvature index has been normalized to $K=2M/r_0^3$ 
\cite{SmollerTemple}. For $K<0$, one finds the solution 
\cite{SmollerTemple} 
\be
\sqrt{a+a^2} -\frac{1}{2} \, \ln \left[ 1+2 \left( a+\sqrt{a+a^2} \right) 
\right] =\sqrt{|K|} \, t 
\ee
with the Big Bang initial condition $a(0)=0$ while, for $K=0$, the scale 
factor and the comoving surface follow the  Einstein-de Sitter  scaling
$ a(t)=\left( \frac{9\pi M}{2} \right)^{1/3} \, t^{2/3} $. 

The surface of the ball, as well as the scale factor of its FLRW dust, 
satisfies the Newtonian energy equation~(\ref{energy}). Indeed Bondi 
\cite{Bondi} remarks the similarity between the relativistic equation for 
the ball expansion $R(t)$ and Eq.~(\ref{energy}), while Mashhoon and 
Partovi \cite{MashhoonPartovi} refer explicitly to the Newtonian analogy 
for FLRW universes (certain errors in \cite{Mansouri, MashhoonPartovi}  
were corrected in \cite{SrivastavaPrasad,Glass}).

\subsection{Vaidya solutions}
\label{subsec:3.2}

If the fluid ball is reduced to a spherical shell and is required to 
expand at the speed of light, the well-known Vaidya solutions 
\cite{Vaidya} apply. The exterior field is still Schwarzschild and matter 
can only be a null dust expanding or contracting at the speed of light. In 
this case, the solution is one of the more well-known Vaidya solutions 
\cite{Vaidya}. The motion of the shell follows a null geodesic. Again, 
this is possible because there is no pressure: the stress-energy tensor of 
a null dust is
\be
T_{ab}=\rho \, k_a k_b \,,
\ee
where $k^a$ is null and geodesic, $k_c k^c=0$ and $k^b \nabla_b k^a=0$.  
This $T_{ab}$ is quite different from the massive perfect fluid 
stress-energy 
tensor because the fluid 4-velocity is now null instead of timelike, but 
the fact that there is no pressure gradient to force the photons away from 
geodesic trajectories survives. Because the shell normal moves along a 
null 
geodesic, the analogy of Sec.~\ref{sec:2} with the Milne universe 
applies again, hence there is a cosmological analogy but the analogous 
universe is the trivial empty one (albeit disguised under the Milne mask).

A ball or a spherical shell expanding at the speed of light will always 
engulf a massive test particle moving radially and starting just above the 
ball. However, a radial outgoing photon emitted above the ball toward 
infinity will not be reached by it (this situation takes us back to the 
previous section).

\section{Ricci and Weyl tensors and quasilocal energy}
\setcounter{equation}{0}
\label{sec:5}

In GR, gravity is curvature and is described by the Riemann 
tensor $R_{abcd}$, which splits into a Ricci part constructed with 
$R_{ab}$ and a Weyl 
part $ C_{abcd}$ \cite{Wald},
\begin{equation}
R_{abcd}=C_{abcd} + g_{a[c}R_{d]b} -g_{b[c} R_{d]a} 
-\frac{ {\cal R} }{3} \, g_{a[c} g_{d]b} \,, 
\end{equation}
where ${\cal R} \equiv {R^c}_c$ is the Ricci scalar.  
Further, the Weyl tensor is decomposed into an electric and  
a magnetic part with respect to a chosen timelike observer. The 
electric part $E_{ab}$  has a  Newtonian analogue, while the magnetic part 
$H_{ab}$  does not \cite{Ellis71} and contains the  true (propagating) 
degrees of freedom of the gravitational field. 

Let the timelike vector $u^a$ be the 4-velocity of an observer: following 
the definition of \cite{Bertschinger} (which differs from that of 
\cite{Ellis71} in the magnetic part of the Weyl tensor, correcting a sign) 
the  electric and  magnetic parts of the 
Weyl tensor are 
\begin{eqnarray}
E_{ac}(u) &=&  C_{abcd} u^b u^d \,,\\
&&\nonumber\\
H_{ac}(u) &=&  \frac{1}{2}\, \eta_{abpq} {C^{pq}}_{ce} u^b 
u^e   \,,
\end{eqnarray}
respectively, where $\eta_{abcd}=\sqrt{-g} \, 
\epsilon_{abcd}$, $\epsilon_{abcd}$ is the Levi-Civita   
symbol and $g$ is the determinant of the metric tensor 
$g_{ab}$, so that  $
\eta^{abcd} = \eta^{[abcd]} $ 
 and  $\eta^{0123}=1/\sqrt{-g} $. According to the observer $u^a$, 
$E_{ab}$ and $H_{ab}$ are purely spatial and symmetric, trace-free 
tensors,
\be
E_{ab}u^a=E_{ab}u^b=
H_{ab}u^a=H_{ab}u^b=0 \,,
\ee
\be
E_{ab}=E_{(ab)} \,, \;\;\;\;\;\;
H_{ab}=H_{(ab)} \,,
\ee
\be
{E^a}_a= {H^a}_a= 0 \,.
\ee
The Weyl tensor is reconstructed from its electric and 
magnetic parts according to \cite{Ellis71, Bertschinger}
\begin{widetext}
\be
C_{abcd}= \left( g_{abef} g_{cdpq} 
-\eta_{abef}\eta_{cdpq}\right) u^e u^p E^{fq} 
-\left( \eta_{abef} g_{cdpq} +g_{abef} \eta_{cdpq} \right) 
u^e u^p H^{fq} 
\ee
with
\be
g_{abef} \equiv g_{ae}  g_{bf} -  g_{af} g_{be} \,,
\ee
giving
\begin{eqnarray}
C_{abcd} &=& u_a u_c E_{bd} - u_a u_d E_{bc} - u_b u_c E_{ad}
+ u_b u_d E_{ac} - \eta_{abef}\eta_{cdpq} u^e u^p E^{fq} \nonumber\\
&&\nonumber\\
&\,& - \eta_{abef} u_c u^e H^{f}_d 
+ \eta_{abef} u^e u_d H^{f}_c
-  u_a u^p   \eta_{cdpq} H^{q}_b
+  u_b u^p   \eta_{cdpq} H^{q}_a \,.
\end{eqnarray}
\end{widetext}
By construction, in the GR extension of the uniform Newtonian ball the 
Riemann tensor is purely Ricci in the interior  and purely Weyl outside. 
In fact, all FLRW metrics are conformally 
flat and the Weyl tensor $ {C_{abc}}^d $ is conformally invariant 
\cite{Wald}, therefore it vanishes in FLRW leaving only the Ricci part of 
$R_{abcd}$ inside the ball. In the exterior 
vacuum region the 
Ricci tensor $R_{ab}=8\pi \left(T_{ab}-g_{ab} T/2\right)$ is 
identically zero, leaving only the Weyl part of $R_{abcd}$. Therefore, 
Ricci and Weyl tensors switch roles when crossing the ball boundary.

At the ball surface, both Ricci and Weyl are 
discontinuous. Focussing on the respective scalars $ {\cal 
R} \equiv {R^a}_a $ and $ C_{abcd} C^{abcd}$, the Ricci scalar $ {\cal 
R}=-\rho+3P =-\rho$ jumps discontinuously to zero at the ball surface, 
while the Weyl scalar is identically zero inside and jumps to the 
Schwarzschild value $C_{abcd}C^{abcd}=48m^2/r^6$.

Since our GR problem originates in Newtonian physics, it is fit to discuss 
the Newtonian character of the (quasilocal) mass used in the discussion.

The interior mass~(\ref{OTmass}) matches the exterior 
Schwarzschild  mass $m$ at the surface of the ball.  For any  $K$, the 
mass $M$ at areal radius $R\leq R_0$ in the FLRW interior is 
\be
M^{(-)}(R)= \frac{4\pi \rho}{3} \, R^3
\ee
and, as noted, it coincides with the Misner-Sharp-Hernandez quasilocal  
energy. In the limit $R\rightarrow a(t)r_0$ to the surface of 
the ball, which is comoving, $M^{(-)}$ becomes constant, and this is 
possible only because the energy density has the dust scaling  
$\rho \sim 1/a^3$, which is necessary to match the constant mass 
$m$ of the Schwarzschild exterior. More generally, the 
Misner-Sharp-Hernandez mass of  a comoving sphere of radius $R$ in any 
FLRW 
space satisfies \cite{myPRDhorizons, mybook}
\be
\dot{M}_\text{MSH}+3H \, \frac{P}{\rho} \, M_\text{MSH}=0 
\ee
and the constancy of $M_\text{MSH}$ goes hand in hand with the vanishing of 
the pressure (it is a peculiarity of the Misner-Sharp-Hernandez mass to 
depend on $\rho$ but not on $P$, while $\dot{M}_\text{MSH}$ depends on $P$ 
but not on $\rho$). 

In principle, a different dependence of the energy density on the scale 
factor $\rho(a)=\rho_0 /a^{3(w+1)}$, which occurs for a fluid equation of 
state $P=w\rho$, $w=$~const., could be compensated if the ball 
boundary expands in the non-comoving way  $R_0(t)\simeq 
t^{w+1}$ to keep $M^{(-)}$ constant.  The Misner-Sharp-Hernandez mass of a 
sphere of radius $R_0$ (not comoving in general) satisfies 
\cite{myPRDhorizons, mybook} 
\be
\dot{M}_\text{MSH}+ 4\pi \rho R_0^3 \left[ H \left(1+\frac{P}{\rho} \right) 
-\frac{ \dot{R}_0}{R_0} \right] = 0 
\ee
and $M_\text{MSH}$ remains constant if $\dot{R}_0/R_0 = (w+1) H$ for 
$P=w\rho$ with $w=$~const. However, this choice would make the pressure 
$P$ discontinuous at the ball boundary, this surface would no longer 
follow a timelike geodesic, and a test particle initially placed on it 
would immediately detach from it. This situation is unacceptable because 
then radial timelike geodesics inside and outside, together with the areal 
radii $R$ and $\bar{r}$, would not match, signalling a discontinuity in 
the geometry. A discontinuity in $P$ is associated with a material layer 
at the ball surface, which becomes a membrane with its own pressure 
foreign to the original Newtonian situation.

As noted, the mass used in the Oppenheimer-Tolman exterior is the 
Misner-Sharp-Hernandez mass common in spherical fluid dynamics \cite{MSH1, 
MSH2}. The more general Hawking-Hayward quasilocal energy 
\cite{Hawking, Hayward} reduces to it in 
spherical symmetry \cite{Haywardspherical} and to the ADM mass 
 if, further, spacetime is asymptotically flat. Several 
other inequivalent quasilocal energy constructions populate the 
literature (see \cite{Szabados} for a review).  In our case, 
when the exterior Oppenheimer-Tolman metric is reduced to Schwarzschild, 
the Hawking-Hayward/Misner-Sharp-Hernandez mass $M_\text{MSH}^{(+)}$ 
reduces to 
the  Schwarzschild mass, {\em i.e.}, the constant $m$ appearing in the 
Schwarzschild metric~(\ref{Schwarzschild}).

In general, the Hawking-Hayward quasilocal mass splits into a contribution 
from the matter stress-energy tensor $T_{ab}$ and a ``pure gravity'' 
contribution from the Weyl tensor. The latter arises solely from the 
electric part of the Weyl tensor, while the magnetic part does not 
contribute \cite{VFSymmetry2015}.

Specifically, the Hawking-Hayward mass enclosed by a 
2-surface ${\cal S}$ is defined as follows
\cite{Hawking, Hayward}. In a  spacetime with metric $ g_{ab}$,  
let ${\cal S}$ be a spacelike, 
embedded, compact, and orientable 2-surface.  $h_{ab}$ and ${\cal 
R}^{(h)}$ denote 
the 2-metric  and Ricci scalar induced on ${\cal S}$ by  $g_{ab}$. Let 
$\mu$ be the volume 
2-form on ${\cal S}$ and $A$ the area  of 
${\cal S}$. The  
congruences of ingoing ($-$) and outgoing ($+$) null  
geodesic emanate from ${\cal S}$ and have expansion scalars   
$\theta_{(\pm)}$ and shear tensors $\sigma_{ab}^{(\pm)}$. 
Let $\omega^a$ denote the 
projection onto ${\cal S}$ of the commutator of 
the null normal vectors to ${\cal S}$, {\em i.e.}, the 
anoholonomicity \cite{Hayward}.  The 
Hawking-Hayward quasilocal mass is \cite{Hawking, 
Hayward}
\begin{eqnarray}
M_\text{HH} & \equiv & \frac{1}{8\pi } \sqrt{ \frac{A}{16\pi}} 
\int_{\cal S} 
\mu \Big( {\cal R}^{(h)} +\theta_{(+)} \theta_{(-)} 
-\frac{1}{2} \, \sigma_{ab}^{(+)} 
\sigma^{ab}_{(-)}   \nonumber\\
&&\nonumber\\
&\, &   -2\omega_a\omega^a \Big) 
\,.\label{HHmass}
\end{eqnarray}
By splitting the Riemann tensor into Ricci and Weyl and, further, the 
latter  into its electric and magnetic parts with respect to an observer 
with 4-velocity parallel to the unit normal to the 2-sphere ${\cal S}$, 
the Hawking mass splits as \cite{VFSymmetry2015} 
\begin{eqnarray}
&& M_\text{HH} =  
\sqrt{ \frac{A}{16\pi}} \int_{{\cal S}}\mu
\left( h^{ab}T_{ab}- \frac{2T}{3}  \right)  \nonumber\\
&&\nonumber\\
&\, & -
\frac{1}{8\pi } \sqrt{ 
\frac{A}{16\pi}} \int_{{\cal S}}\mu \,
\eta_{abef}\eta_{cdpq}h^{ac} h^{bd}u^e u^p E^{fq}
\,,  \label{summary}
\end{eqnarray}
where the ``pure gravity'' contribution to $M_\text{HH}$ (the second 
integral  
in the right-hand side of Eq.~(\ref{summary})) comes only from the 
electric part $E_{ab}$ of the Weyl tensor, while the magnetic part 
$H_{ab}$ does not contribute. In this sense, the Hawking mass is 
``Newtonian''.

In the spherical spacetime corresponding to the Newtonian ball, the 
2-surface $ {\cal S}$ is a sphere of radius $R$ or $\bar{r}$ and 
Eq.~(\ref{summary}) reduces to  
\be
M_\text{MSH}(R)= \frac{4\pi R^3}{3} \, \rho \, \theta_H\left( R_0-R\right) 
+m\,  
\theta_H\left( \bar{r}-\bar{r}_0\right) \,,
\ee
where $R$ stands for the areal radius ($R=a(t)r$ inside and $\bar{r}$ 
outside) and $\theta_H (x)$ is the Heaviside step function.  
In the interior of the ball, $M_\text{MSH}$ is given entirely by the first 
term on the right-hand side, while the second term vanishes. In the 
exterior region, these two terms switch roles. The matching between 
interior and exterior makes the two terms equal on the ball boundary and 
guarantees the continuity of $M_\text{MSH}$. At all times, this quasilocal 
energy remains ``Newtonian''.\footnote{A Newtonian character for the 
Misner-Sharp-Hernandez quasilocal mass is claimed also in 
Ref.~\cite{BlauRollier} based on the study of the timelike radial 
geodesics of 
Schwarzschild.} Ultimately, the fact that a Newtonian 
analogy exists for FLRW cosmology is made possible by the fact that the 
corresponding GR manifold has  vanishing magnetic part of the Weyl 
tensor everywhere according to static observers.

\section{Back to Newtonian gravity from FLRW: symmetries}
\setcounter{equation}{0}
\label{sec:6}

Symmetries of the Einstein-Friedmann equations have been explored in 
detail, mostly with the goal of generating new exact 
solutions ({\em e.g.}, 
\cite{ParsonsBarrow,Chimento,FaraoniPLB,BarrowPaliathanasis, 
VFSymmetry2020,Pailasetal20}). 
For FLRW universes fueled by a perfect fluid, in general one can rescale 
appropriately time and scale factor or Hubble parameter, while changing to 
a different barotropic fluid, leaving the Einstein-Friedmann equations 
unchanged. Here we are concerned, in particular, with 
the symmetry found in Ref.~\cite{Chimento}  for spatially 
flat ($K=0$) FLRW universes\footnote{The  fact that this symmetry also 
relates de Sitter 
and anti-de Sitter spaces apparently went unnoticed in the literature.}
\begin{eqnarray}
\rho & \rightarrow &  \bar{\rho}=\bar{\rho} (\rho) \,,\label{symm1}\\
&&\nonumber\\ 
H & \rightarrow &  \bar{H}=\sqrt{ \frac{ \bar{\rho}}{\rho}  }  \, \, H 
\,,\label{symm2}\\
&& \nonumber\\
P & \rightarrow &  \bar{P }=-\bar{\rho} + \sqrt{ \frac{ \rho}{\bar{\rho}} 
}  \, 
\left( P+\rho\right) \, \frac{ d\bar{\rho} }{d\rho}  \label{symm3}  
\end{eqnarray}
(where $\bar{\rho}(\rho)$ is  a free but positive and differentiable 
function), 
which leaves the Einstein-Friedmann equations invariant. Since the 
relativistic analogue of the Newtonian ball requires the interior FLRW 
metric to match the exterior Schwarzschild and, therefore, the pressure 
$P$ to vanish everywhere, in order for this transformation to be a 
symmetry of the Newtonian analogue, it must be $P=\bar{P}=0$. In this 
case Eq.~(\ref{symm3}) becomes, using the new variable $z \equiv 
1/\sqrt{\rho}$,
\be
\frac{d\bar{z}}{dz}=\frac{ \bar{\rho}^{-3/2} }{\rho^{-3/2}}\, 
\frac{d\bar{\rho}}{d\rho}=1 
\ee
and is trivially integrated to $\bar{z}(z)=z-z_0$, or
\be
\frac{ \bar{\rho}}{\rho} = \frac{1}{ 1+ \frac{\rho}{\rho_0} -2\sqrt{ 
\frac{\rho}{\rho_0}} } \,,
\ee
where $\rho_0$ (or $z_0$) is an integration constant. Therefore, the 
particular transformation
\begin{eqnarray}
\rho & \rightarrow &  \bar{\rho}=\frac{  \rho}{1+ \frac{\rho}{\rho_0} -2\sqrt{ 
\frac{\rho}{\rho_0} } }   \,,\label{S1}\\
&&\nonumber\\ 
H & \rightarrow &  \bar{H}=\frac{H}{  1+ \frac{\rho}{\rho_0} -2\sqrt{ 
\frac{\rho}{\rho_0}} }  \,,\label{S2}   
\end{eqnarray}
preserves the zero-pressure condition and does not change the 
physics of the Newtonian ball (Eq.~(\ref{energy}) with $E=0$)  
used in the cosmological analogy. Since we have $K=0$ and a dust, 
$ R(t)=R_0 t^{2/3} $ and one obtains
\begin{eqnarray}
\bar{ \rho} (t) &=& \frac{3M}{4\pi  R_0^3} \, 
\frac{1}{ 
t^2 -2\sqrt{ \frac{3M}{4\pi R_0^2 \rho_0} } \, t 
+\frac{3M}{4\pi R_0^3 \rho_0}  
} \,,\\
&&\nonumber\\
\overline{ \left( \frac{ \dot{ R }}{ R} \right) } &=& 
\left( \frac{\dot{R}}{R}  \right) 
\frac{ t^2}{ t^2 -2 \sqrt{\frac{3M}{4\pi R_0^3 \rho_0}} \, t 
+\frac{3M}{4\pi R_0^2 \rho_0} } \,.
\end{eqnarray}
In spite of the fact that 
uniform balls are often used in teaching Newtonian physics and in 
well-known problems such as gravity tunnels 
\cite{Routh, Cooper, Kirmser, Venezian, Mallett, Laslett} and the 
terrestrial brachistochrone \cite{Cooper, Venezian, new} (also discussed  
in view of futuristic technological applications 
\cite{app}), this symmetry seems to have escaped attention.

Finally, another aspect of the Einstein-Friedmann equations that has been 
used extensively in FLRW cosmology is the fact that, in the presence of a 
single barotropic fluid with a constant equation of state $P=w\rho$, the 
Friedmann equation can be reduced to a Riccati equation by changing 
comoving into conformal time (\cite{Jantzen,myAJP,Rosu}).  This means 
that, applying the change of variable for a dust-dominated FLRW universe 
to the Newtonian equation of motion of a test particle, the same reduction 
occurs. Indeed, the change
\begin{eqnarray}
r &=& s^2 \,, \label{C1}\\
&&\nonumber\\
dt &=& r d\eta = s^2 d\eta \label{C2}\,,
\end{eqnarray}
(where the last equation is analogous to the change from comoving to 
conformal time $dt=ad\eta$ of FLRW cosmology) reduces the Newtonian 
equation of motion
\be
\frac{d^2 \vec{r} }{dt^2}= -\frac{GM}{r^3} \, \vec{r}
\ee
to
\be
\frac{d^2 \vec{s}}{d\eta^2}+\frac{|E|}{2}\, \vec{s}=0 \,,
\ee
where $E = \frac{1}{2} \left( d \vec{r}/dt \right)^2 -\frac{GM}{r}<0$  is 
the particle energy. The change of variables (\ref{C1}), (\ref{C2}) reduces 
the Coulomb force problem to that of two decoupled  harmonic oscillators. 
This change of 
variables for the Newtonian problem has been known since Euler and has 
been discussed or rediscovered many times through the years \cite{Euler}.

\section{Discussion and conclusions}
\setcounter{equation}{0}
\label{sec:7}

Abandoning the original pedagogical motivation for the Newtonian analogy 
to FLRW universes, one turns it into a rather interesting fully 
relativistic problem. It is 
natural, with a slight deviation from the original theme, to consider as 
the first model the radial geodesic trajectories of massive 
particles 
starting out just above the ball surface. This is very easy to do, but it 
has 
the drawback that the ball surface could meet and engulf the test 
particle. A formal analogy between the radial timelike geodesic equation 
and the Friedmann equation ensues, and the analogous FLRW universe can 
only be dust-dominated.

As a second model, we have considered radial null geodesics described by 
outgoing photons starting just above the ball surface. These will never be 
engulfed by the ball and they also give rise to a formal analogy that is 
not, however, very interesting because it reproduces only the empty 
Milne universe. More general static and spherically symmetric spacetimes 
generate new cosmological analogies through their timelike and null radial 
geodesics: we have listed several of them in the Appendix.

Finally, the full general relativistic problem of a uniform fluid ball 
expanding (or contracting) in a surrounding Schwarzschild vacuum can be 
considered. This situation is a special case of more general set-ups in 
the literature on fluid spheres expanding in a (possibly cosmological) 
fluid. One quickly learns that two assumptions of the Newtonian analogy, 
usually not explained in the introductory literature,  are crucial: {\em 
i)}~{\it Uniform ball.} A non-uniform ball leads to a spherical 
inhomogeneous universe, which is not an analogy for FLRW. {\em ii)}~{\it 
Test fluid.} As a consequence of the Darmois-Israel matching 
conditions, only a dust interior can match the Schwarzschild exterior. 
Moreover, a test particle initially placed on the surface of a ball of 
fluid other than dust would immediately detach from it, compromising the 
analogy. For 
a massive test particle to remain on the surface of the ball, the fluid in 
it must be dust.

The pressure must vanish at the surface of the ball to avoid a matter  
layer on it, while it is accepted as a necessary evil that the density is 
discontinuous there, as in the Schwarzschild interior solution.

To broaden the scope, one can in principle consider other physical 
situations. For example, one can study a discontinuous pressure associated 
with a spherical matter layer enclosing a fluid with $P\neq 0$. At this 
point, however, there is no longer a need for the ball and one can retain 
only the spherical shell since, due to the Birkhoff theorem, the exterior 
geometry is still Schwarzschild. Now the problem bears little resemblance 
to an expanding universe mimicked by a uniform ball.

Inhomogeneous universes analogous to non-uniform balls with 
$\rho=\rho(t,r), P=P(t,r)$ \cite{Krasinskibook} also do not resemble FLRW 
ones.  In principle, 
one could also consider theories of gravity alternative to GR to evade the 
Birkhoff theorem and match an interior ball solution to an exterior 
geometry that is not Schwarzschild. All these options are found wanting 
because of one crucial point: the Schwarzschild solution is the unique 
solution of the vacuum Einstein equations that is spherically symmetric 
and asymptotically flat. In all the alternatives mentioned above, instead, 
there is no unique solution to the field equations, hence these models are 
not as compelling. Looking for such alternatives, one goes further and 
further away from the simple Newtonian analogy, creating progressively 
more complicated and physically unjustified situations. To conclude,  
extensions to GR of the heuristic Newtonian ball problem have reasonably 
straightforward solutions and do create new formal analogies with 
FLRW universes.

\begin{acknowledgments} 
This work is supported, in part, by the Natural 
Sciences \& Engineering Research Council of Canada (Grant no. 2016-03803 
to V.F.) and by Bishop's University.
\end{acknowledgments}

\appendix
\section{Examples of analogies from radial geodesics of static 
spherical geometries}
\renewcommand{\theequation}{A.\arabic{equation}}

Here we provide examples of cosmic analogies arising from the radial 
geodesics of well-known static and spherically symmetric metrics, 
beginning with situations in which $\Phi\equiv 0$ and $g_{tt}\,g_{RR}=-1$.

\subsection{Reissner–Nordstr\"om metric}

The first obvious candidate is the Reissner-Nordtr\"om metric
\begin{equation}
ds^2 = -\left( 1- \frac{2m}{R} + \frac{Q^2}{R^2} \right) dt^2 + 
\frac{dR^2}{ 1-2 
\frac{m}{R} + \frac{Q^2}{R^2} } + R^2 d\Omega_{(2)} ^2 \,,
\end{equation}
where the constants $m$ and $Q$
 are the mass and charge parameters. The Misner-Sharp-Hernandez mass is  
$
M(R) = m -Q^2/(2R) $ 
and the analogous energy density in the Friedmann equation is 
\begin{equation}
\rho (a) = \frac{3}{4\pi} \,\frac{m}{a^3} - \frac{3}{8\pi}\, 
\frac{Q^2}{a^4}\,.
\end{equation}
The second term on the right-hand side corresponds to a radiation fluid 
with an unphysical negative energy density, which spoils the  
analogy. However, when $Q=0$ the Reissner-Nordstr\"om metric reduces to  
Schwarzschild and the energy density coincides with the first term, a 
positive 
dust density $\rho = \rho_{0}/a^3$ for a universe with any possible 
sign of the curvature index, as seen in Sec.~\ref{subsec:1.2}.  

\subsection{(Anti-)de Sitter space}

The (Anti-)de Sitter line element in locally static coordinates  is
\begin{equation}
ds^2 = - \left( 1 \mp H^2 R^2 \right) dt^2 + \frac{dR^2}{1 \mp H^2 R^2)} 
+ R^2 d\Omega_{(2)}^2
\end{equation}
where the upper (resp. lower) sign refers to de Sitter (resp, Anti-de 
Sitter) space. The Misner-Sharp--Hernandez mass is $ M(R) = \pm H^2 R^3/2 
$ and the energy density and pressure of the analogous FLRW universe are
\begin{equation}
\rho(a) =  \pm \frac{3 H^2}{8 \pi} \,,
\end{equation}
\begin{equation}
P = - \frac{M'}{4\pi a^2} = \mp \frac{3 H^2 a^2}{8\pi a^2} =- \rho \,,
\end{equation}
that is, the energy density and pressure of a cosmological 
constant $\Lambda =\pm 3H^2$. Thus, the timelike geodesics of 
(Anti-)de Sitter produce analogous universes with the same cosmological 
constant and any value of the curvature index. For $K=0, -1$, 
respectively, this 
procedure reproduces  the same (Anti-)de Sitter space used as an 
analogue generator. 

\subsection{Schwarzschild-de Sitter/Kottler geometry}

The Schwarzschild-de Sitter/Kottler line element 
\begin{eqnarray}
ds^2 &=& - \left( 1- \frac{2m}{R} - H^2 R^2 \right)dt^2 + 
\frac{dR^2}{1-\frac{2m}{R} - H^2 R^2} \nonumber\\
&&\nonumber\\
&\, & + R^2 d\Omega_{(2)}^2
\end{eqnarray}
has Misner-Sharp-Hernandez mass $M = m +H^2 R^3/2 $, generating the 
 energy density of the FLRW analogue cosmic fluid
\begin{equation}
\rho (a) = \frac{3m}{4  \pi a^3} + \frac{3H^2}{8\pi} \,,
\end{equation}
corresponding to a cosmological constant plus a dust. This was to be 
expected since the Schwarzschild black hole generates a dust-dominated 
analogous FLRW universe, while de Sitter space has itself as an analogue.

\subsection{Kiselev solution}

The Kiselev line element \cite{Kiselev} describes a black hole embedded 
in a mixture of  fluids with anisotropic pressure (contrary to 
appearances, 
it is not asymptotically FLRW nor  a perfect fluid solution 
\cite{Visser}), and it reads
\begin{equation}
ds^2 = - f(R) dt^2 + \frac{dR^2}{f(R)} + R^2 d\Omega_{(2)}^2 \,,
\end{equation}
where 
\be
f(R) = 1 - \frac{2m}{R} - \sum_n \left( \frac{r_n}{R} \right)^{3w_{n} +1} 
\,,
\ee
where $m, r_n$, and $w_n$ are constants and $-1 < w_n <-1/3 $ 
\cite{Kiselev}. Here the Misner-Sharp-Hernandez mass is 
\begin{equation}
M(R) = m + \frac{R}{2} \sum_n \left( \frac{r_n}{R} \right)^{3w_n+1}\,,
\end{equation}
producing the  energy density and pressure  of the analogous FLRW universe 
\begin{eqnarray}
\rho &=& \rho_\text{dust} +\sum_n  
\frac{3a_n^{3w_n +1}}{8 \pi} \, \frac{1}{a^{3\left( w_n +1 \right)}}\,,\\
&&\nonumber\\
P &=& \sum_n w_n \rho_{n}  \,,
\end{eqnarray}
where 
\begin{eqnarray}
\rho_\text{dust} &=& \frac{\rho_{0} }{a^3} \,,\\ 
&&\nonumber\\
\rho_{0} &=& \frac{3}{4 \pi} \,,\\
&&\nonumber\\
\rho_{n} &=& \frac{3a_n^{3w_n +1} }{8 \pi} \,.
\end{eqnarray}
In this analogy, the Kiselev anisotropic fluid becomes a mixture of   
(isotropic) perfect quintessence fluids.

\subsection{Barriola-Vilenkin global monopole}

The Barriola-Vilenkin global monopole (or ``string hedgehog'') has the 
geometry \cite{BarriolaVilenkin}
\begin{equation}
ds^2 = - \left( 1-8\pi \eta^2 - \frac{2m}{R} \right)dt^2  + 
\frac{dr^2}{1-8\pi \eta^2 - \frac{2m}{R}} + R^2 d\Omega_{(2)}^2 
\end{equation}
with constant $m, \eta$ and the Misner-Sharp-Hernandez mass is $M(R) = m + 
4\pi \eta^2 \, R$,  
producing the energy density 
\begin{equation}
 \rho(a) = \frac{3m}{4\pi a^3} + \frac{3 \eta^2 }{a^2}
\end{equation}
and the analogous Friedmann equation
\begin{equation}
H^2 = \frac{\left( \bar{E}^2 -1 +3 \eta^2\right)}{a^2} + 
\frac{\rho_{0}}{a^3}
\end{equation}
with  
\begin{eqnarray}
K &=& 1-\bar{E}^2 - 3\eta^2\,,\\
&&\nonumber\\
\rho_{0} &=& \frac{3m}{4\pi} \,;
\end{eqnarray}
this is a spatially curved  universe (except for special values of 
$\bar{E}$ and $\eta$) filled with dust.

\subsection{Bardeen regular black hole}

The Bardeen regular black hole \cite{Bardeen} quantum-corrects the 
Schwarzschild black hole to remoce the central singularity and it solves  
the Einstein equations coupled to nonlinear electrodynamics 
\cite{nonlinearED}. The line element 
\begin{equation}
ds^2 = -\left[1- \frac{2m R^2}{\left( R^2 +\alpha^2\right)^{3/2} } 
\right] dt^2 + 
\frac{dR^2}{1- \frac{2mR^2}{\left(R^2 +\alpha^2\right)^{3/2} }} + 
R^2 d\Omega_{(2)}^2 
\end{equation}
 gives the 
Misner-Sharp-Hernandez mass \begin{equation}
M(R) = \frac{mR^3}{\left( R^2 +\alpha^2 \right)^{3/2} }
\end{equation}
corresponding to the energy density and pressure of the analogous FLRW 
universe
\begin{eqnarray}
\rho(a) &=& \frac{3m}{4\pi} \, \frac{1}{\left( a^2 +a_0^2 
\right)^{3/2} } \,,\\
&&\nonumber\\
P &=&  - \left( \frac{4\pi}{3m} \right)^{2/3} \alpha^2  \rho^{5/3} \,,
\end{eqnarray}
satisfying a phantom and nonlinear equation of state.

Let us move now to static spherical geometries in which  $\Phi \neq 
0$.

\subsection{Morris-Thorne wormhole}

Let us consider the Morris-Thorme wormhole \cite{MT1} with line element 
\begin{equation}
ds^2 = -dt^2 +dr^2 + \left( b_0 ^2 + r^2 \right) d\Omega_{(2)}^2 
\label{MT1}
\end{equation}
where $b_0$ is a constant and the areal radius is $ R = \sqrt{b_0 ^2 + 
r^2}$. Substituting  $ dr =RdR/ r $ into Eq.~(\ref{MT1}) gives
\begin{equation}
	ds^2 = - dt^2 + \frac{R^2}{R^2 -b_0^2}\, dR^2 + R^2 
d\Omega_{(2)}^2 \,.
\end{equation} 
The Misner-Sharp-Hernandez mass is $
M(R) = b_0 ^2/(2R)$, $ \mbox{e}^{-2\Phi} = \left( 1 - 2m/R 
\right)^{-1}$, and the energy density of the FLRW cosmic analogue is  
\begin{equation}
 	\rho(a) = \frac{3b_0^2}{8\pi a^4} \,. 
\end{equation}
The Friedmann equation satisfied by the analogous FLRW universe reads 
\begin{equation}
H^2 = \frac{\left(E^2 - 1\right)}{a^2} +  \frac{b_0^2\left( 
1-E^2\right)}{a^4} \,,
\end{equation} 
which makes sense physically if  $E^2 <1$ and negative 
energy densities are avoided; then the analogous universe is spatially 
curved and contains a radiation fluid.

Radial null geodesics produce the analogous Friedmann equation
\be
 H^2 = \frac{E^2}{a^2} - \frac{b_0^2 E^2 }{a^3}
\,
\ee
with a dust of negative energy density, which is unphysical.

\subsection{Wyman's ``other'' solution}

Wyman's ``other'' solution \cite{Wymanother} is a little known scalar 
field solution of the Einstein equations, not to be confused with the 
better known solution (re-)discovered by Fisher, Bergmann \& 
Leipnik, Janis, Newman \& Winicour, Buchdahl, and Wyman \cite{FBLJNWBW, 
Wymanother}. The line element is
\begin{equation}
ds^2 = -R^2dt^2 +2 dR^2 + R^2 d\Omega_{(2)}^2 \label{Wother}
\end{equation}
with time-dependent scalar field source $\phi (t) = \phi_0 t$. Here 
$
\mbox{e}^{-2\Phi} = R^2 /\left(1-2M/R \right) $, $ M(R)= R/4$, 
and the analogous Friedmann equation is 
\begin{equation}
H^2 = \frac{8 \pi \rho_{0}}{3 a^4} - \frac{K}{a^2}
\end{equation}
where $K=-1/2$. This analogous universe is positively curved and filled 
with radiation. 
 
Radial null geodesics, instead, produce the analogous Friedmann equation $ 
H^2 = E^2/(2 a^4) $ describing a spatially flat, radiation-dominated FLRW 
universe.

The geometry~(\ref{Wother}) has been generalized by including a 
cosmological constant $\Lambda$ \cite{Sultana} and studied in 
\cite{Behnaz}:
\begin{equation}
ds^2 = -R^2dt^2 + \frac{2 dR^2}{1-2\Lambda R^2 /3} + R^2 d\Omega_{(2)}^2 
\,.
\end{equation}
Looking at radial timelike geodesics, the Misner-Sharp-Hernandez mass is 
now 
$M(R)=\frac{R}{4}+\frac{\Lambda 
R^3}{6}$ and the analogous Friedmann equation becomes
\be
H^2 = \frac{ \bar{E}^2 \left(1+\Lambda/3\right) -1/2}{a^2} 
+\frac{\Lambda}{3} +\frac{ \bar{E}^2}{2a^4} \,,
\ee
corresponding to a FLRW universe with curvature index $K=\frac{1}{2} 
-\bar{E}^2 \left( 1+\Lambda/3 \right) $, cosmological constant $\Lambda$, 
and a radiation fluid with $\rho_0= \bar{E}^2/2$. The analogy stemming 
from radial null geodesics is unchanged.


\begin{thebibliography}{99}

\bibitem{Ryden} B. Ryden, {\em Introduction to Cosmology} (Addison Wesley, 
San Francisco, 2003).

\bibitem{Liddle} A. Liddle, {\em An Introduction to Modern Cosmology} 
(Wiley, New York, 2015). 

\bibitem{Inverno} R. d'Inverno, {\em Introducing Einstein's General 
Relativity} (Clarendon Press, Oxford, 1998).

\bibitem{Slava} V. Mukhanov, {\em Physical Foundations of Cosmology} 
(Cambridge University Press, Cambridge, 2005).

\bibitem{Harrison} E.R. Harrison,  
{\em Ann. Phys. (NY)} {\bf 35}, 437 (1965).

\bibitem{Michell} J. Michell,  
{\em Phil. Trans. Roy. Soc.} {\bf 74}, 35 (1784).

\bibitem{Laplace} S. Laplace, 
{\em Allgemeine Geographische Ephemeriden} {\bf 4}, 1 (1799). 
	Translated in S.W. Hawking and G.F.R. Ellis, {\em The Large Scale 
Structure of Space-Time} (Cambridge University Press, Cambridge, 1973), 
pp.~365-368.

\bibitem{Lanczos} C. Lanczos, {\em Ann. Phys. (Leipzig)} {\bf 379}, 518 
(1924).

\bibitem{Darmois} G. Darmois, {\em Memorial des Sciences Mathematiques}  
XXV (Gauthier-Villars, Paris, 1927).

\bibitem{Israel} W. Israel, {\em Nuovo Cimento B} {\bf 44}, 1 (1966); {\em 
Errata} {\bf 48}, 463(E) (1967).

\bibitem{OppenheimerSnyder} J.R. Oppenheimer and J.R. Snyder, 
{\em Phys. Rev.} {\bf 56}, 455 (1939).

\bibitem{Wald} R.M. Wald, {\em General Relativity} (Chicago University 
Press, Chicago, 1984).

\bibitem{Vieira1} H.S. Vieira and V.B. Bezerra, 
{\em J. Math. Phys.} {\bf 56}, 092501 (2015).

\bibitem{Vieira2} H.S. Vieira, V.B. Bezerra, C.R. Muniz, and M.S. Cunha, 
{\em J. Math. Phys.} {\bf 60}, 102301 (2019).

\bibitem{NielsenVisser} A.B. Nielsen and M. Visser, {\em Class. Quantum 
Grav.} {\bf 23}, 4637 (2006).

\bibitem{AbreuVisser} G. Abreu and M. Visser, {\em Phys. Rev. D} {\bf 82}, 
044027 (2010).

\bibitem{MSH1} C.W. Misner and D.H. Sharp, {\em Phys. Rev.} {\bf 136}, 
B571 (1964).

\bibitem{MSH2} W.C. Hernandez and C.W. Misner, {\em Astrophys. J.} {\bf 
143}, 452 (1966).

\bibitem{Hawking} S. Hawking, {\em J. Math. Phys.} {\bf 9}, 598 (1968).

\bibitem{Hayward}S.A. Hayward, {\em Phys. Rev. D} {\bf 49}, 831 (1994).

\bibitem{Haywardspherical} S.A. Hayward, {\em Phys. Rev. D} {\bf 53}, 1938 
(1996).

\bibitem{Tolman} R.C. Tolman, {\em Relativity, Thermodynamics and 
Cosmology} (Oxford University Press, London, 1934).

\bibitem{Vaidyaball} P.C. Vaidya, 
{\em Phys. Rev.} {\bf 174},  1615 (1968).

\bibitem{SmollerTemple} J. Smoller and B. Temple, 
{\em SIAM J. Appl. Math.} {\bf 58}, 15 (1998).

\bibitem{MashhoonPartovi} B. Mashhoon and M.H. Partovi, {\em Ann. Phys. 
(NY)} {\bf 130}, 99 (1980).

\bibitem{SrivastavaPrasad} D.C. Srivastava and S.S. Prasad, 
{\em Gen. Relativ. Gravit.} {\bf 15}, 65 (1983).

\bibitem{ThompsonWhitrow} A.H. Thompson and W.J. Whitrow, {\em Mon. Not. 
R. Astron. Soc.} {\bf 136}, 207 (1967).

\bibitem{ThompsonWhitrow2} A.H. Thompson and W.J. Whitrow, {\em Mon. Not. 
R. Astron. Soc.} {\bf 139}, 499 (1968).

\bibitem{Bondi} H. Bondi, {\em Mon. Not. R. Astron. Soc.} {\bf 142}, 333 
(1969).

\bibitem{BondiNature} H. Bondi, {\em Nature} {\bf 215}, 838 (1967).

\bibitem{McVittie} G.C. McVittie, {\em Astrophys. J.} {\bf 143}, 682 
(1966).

\bibitem{Mansouri} R. Mansouri, {\em Ann. Inst. Henri Poincar\'e}, {\bf 
A27}, 175 (1977).

\bibitem{Glass} E.N. Glass, {\em J. Math. Phys.} {\bf 20}, 1508 (1979). 

\bibitem{Vaidya} P.C. Vaidya, 
{\em Current Science} {\bf 12}, 183 (1943; 
reprinted 
in {\em Gen. Relativ. Gravit.} {\bf 31}, 119 (1999).

\bibitem{Ellis71} G.F.R. Ellis, ``General Relativity and Cosmology'', in 
{\em Proceedings of the International School of Physics E. Fermi, Course 
XLVII}, Varenna, Italy, 1969, edited by R.K. Sachs (Academic Press, New 
York, 1971), pp.~104-182.

\bibitem{Bertschinger} E. Bertschinger, 
arXiv:astro-ph/9503125.

\bibitem{myPRDhorizons} V. Faraoni, 
{\em Phys. Rev. D} {\bf 84}, 024003 (2011).

\bibitem{mybook} V. Faraoni, {\em Cosmological and Black Hole Apparent 
Horizons} (Springer, New York, 2015).

\bibitem{Jacobson} T. Jacobson, {\em Class. Quantum Grav.} {\bf 24}, 5717 
(2007).

\bibitem{Szabados} L.B. Szabados, 
{\em Living Rev. Relativ.} {\bf 12}, 4 (2009).

\bibitem{VFSymmetry2015} V. Faraoni, 
{\em Symmetry} {\bf 7}, 2038 (2015).

\bibitem{BlauRollier} M. Blau and B. Rollier, 
{\em Class. Quantum Grav.} 25 (2008). 

\bibitem{ParsonsBarrow} P. Parsons and J.D. Barrow, {\em Class. Quantum 
Grav.} {\bf 12}, 1715 (1995).

\bibitem{Chimento} L.P. Chimento, {\em Phys. Rev. D} {\bf 65}, 063517 
(2002).

\bibitem{FaraoniPLB} V. Faraoni, 
{\em Phys. Lett. B} {\bf 703}, 228 (2011).

\bibitem{BarrowPaliathanasis} J.D. Barrow and A. Paliathanasis, {\em Gen. 
Rel. Gravit.} {\bf 50}, 82 (2018).

\bibitem{VFSymmetry2020} V. Faraoni, {\em Symmetry} {\bf 12}, 147 (2020).

\bibitem{Pailasetal20} T. Pailas, N. Dimakis, A. Paliathanasis, 
P.A. Terzis, and T. Christodoulakis, 
arXiv:2005.11726.

\bibitem{Routh} E.J. Routh, {\em A Treatise on Dynamics of a Particle} 
(Cambridge University Press, Cambridge, 1898).

\bibitem{Cooper}P.W. Cooper, 
{\em Am. J. Phys.} {\bf 34}, 68 (1966).

\bibitem{Kirmser}P.G. Kirmser,  
{\em Am. J. Phys.} {\bf 4}, 701 (1966).

\bibitem{Venezian}G. Venezian, 
{\em Am. J. Phys.} {\bf 4}, 701 ( 1966). 

\bibitem{Mallett} R.L. Mallett, 
{\em Am. J. Phys.} {\bf 34}, 702 (1966).

\bibitem{Laslett}L.J. Laslett, 
{\em Am. J. Phys.}  {\bf 34}, 702 (1966).  

\bibitem{new} A.R. Klotz, 
arXiv:1505.05894; 
 W. Dean Pesnella, 
{\em Am. J. Phys.}  {\bf 84}, 192 (2016);  
T. Concannon and G. Giordano, 
arXiv:1606.01852; 
R. Antonelli and A.R. Klotz,  
{\em Am. J. Phys.} {\bf 85}, 469 (2017);  
M. Selmkea, 
{\em Am. J. Phys.} {\bf 86}, 153 (2018); 
W. Dean Pesnella, 
{\em Am. J. Phys.}  {\bf 86}, 338 (2018);
R. Taillet, 
{\em Am. J. Phys.}  {\bf 86}, 924 (2018);   
 M.A. De Andrade and L.G. Ferreira Filho, 
{\em Rev. Bras. Ensino F\'is.} {\bf  
40}, 3 (2018); 
S. Isermann, 
{\em Am. J. Phys.} {\bf 87}, 10 (2019);
S. Isermann, 
{\em Am. J. Phys.} {\bf 87}, 646 ( 2019);
E. Parker, 
{\em Gen. Relativ. Gravit.} {\bf 49}, 106 (2017);
 M. Seel, 
{\em Eur. J. Phys.} {\bf 39}, 3 (2019); 
A. Gjerl\o{o}v and W. Dean Pesnella, 
{\em Am. J. Phys.} {\bf 87}, 452 (2019);
A. Simoni\u{c}, 
arXiv:2001.03279; 
 M. Dragoni, 
{\em The Physics Teacher} {\bf 58}, 97 (2020);

\bibitem{app}M.R. Feldman, J.D. Anderson, G. Schubert, V. Trimble, S.M. 
Kopeikin, and C. L\"ammerzahl, 
{\em Class. Quantum Grav. } {\bf 33}, 125013 (2016); 
H. Xie, J.W. Zhao, H.W. Zhou, S.H. Ren, and R.X. Zhang, 
{\em Tunnelling and Underground Space Technology} {\bf 96}, 103129 (2020); 
W. Hao, Z. Wu, H. Xia, Y. Bo, G. Xin, and C. Ling, 
{\em Thermal Science} {\bf 23}, 1441 (2019).

\bibitem{Jantzen}R.T. Jantzen and C. Uggla, 
{\em Gen. Relativ. Gravit.} {\bf 24}, 59 
(1992).

\bibitem{myAJP}V. Faraoni, 
{\em Am. J. Phys.} {\bf 67}, 732 (1999).

\bibitem{Rosu}M. Nowakowski and H.C. Rosu, 
{\em Phys. Rev. E} {\bf 65}, 047602 (2002).

\bibitem{Euler}L. Euler, {\em Novi Comm. Acad. Sci. Petrop.} {\bf 11}, 144 
(1765);
K. Bohlin, {\em Bull. Astron.} {\bf 28}, 113 (1911); 
K. Sundman, {\em Acta Math.} {\bf 36}, 105 (1912);
T. Levi-Civita, 
{\em Acta Math.} {\bf 42}, 99 (1920);
V.R. Bond, 
{\em Celestial Mechanics} {\bf 35}, 1 (1985);
 J.D. Barrow, 
{\em The Observatory} {\bf 113}, 210 (1993);
 D. Heggie and P. Hut, {\em The Gravitational 
Million-Body Problem} (Cambridge University Press, Cambridge, 2003), pp. 
143–149 (2003).

\bibitem{Krasinskibook} A. Krasi\'{n}ski, {\em Inomogeneous Cosmological 
Models} (Cambridge University Press, Cambridge, 1997).

\bibitem{Kiselev} V.V. Kiselev, {\em Class. Quantum Grav.} {\bf 20}, 1187 
(2003).

\bibitem{Visser} M. Visser, 
arXiv:1908.11058.

\bibitem{BarriolaVilenkin} M. Barriola and A. Vilenkin, {\em Phys. Rev. 
Lett.} {\bf 63}, 341 (1989); E.I. Guendelman and A. Rabinowitz, {\em Phys. 
Rev. D} {\bf 44}, 3152 (1991).

\bibitem{Bardeen} J.M. Bardeen,  ``Non-singular general-relativistic 
gravitational collapse'',  in {\em Proceedings of International
Conference GR5}, Tbilisi, USSR 1968, p.~174.

\bibitem{nonlinearED} E. Ay\'on-Beato and  A. Garc\'ia,
{\em Phys. Lett. B} {\bf 493}, 149 (2000)

\bibitem{MT1} M.S. Morris and K.S. Thorne, 
{\em Am. J. Phys.}  {\bf 56}, 395 (1988).

\bibitem{Wymanother} M. Wyman, {\em Phys. Rev. D} {\bf 24}, 839 (1981).

\bibitem{FBLJNWBW} I.Z. Fisher, {\em Zh. Eksp. Teor. Fiz.} {\bf 18}, 636 
(1948); O. Bergmann and R. Leipnik, {\em Phys. Rev.} {\bf 107}, 1157
(1957);  A.I. Janis, E. T. Newman, and J. Winicour, {\em Phys. Rev.
Lett.} {\bf 20}, 878 (1968); H. A. Buchdahl, {\em Int. J. Theor. Phys.} 
{\bf 6}, 407 (1972). 

\bibitem{Sultana} J. Sultana, {\em Gen. Relativ. Gravit.} {\bf 47}, 73 
(2015).

\bibitem{Behnaz} A. Banijamali, B. Fazlpour, and V. Faraoni, {\em Phys. 
Rev. D} {\bf 100}, 064017 (2019).

\end{thebibliography}

\end{document}